\documentclass[a4paper,12pt]{article}
 \UseRawInputEncoding
\usepackage[left=2.5cm,bottom=3cm,right=2.5cm,top=3cm]{geometry} 
 \usepackage{mathtools}  
\usepackage{makecell}   %
\usepackage{bigints}
\usepackage{overpic,youngtab}
\usepackage{subfigure}
\usepackage{bbm}
\usepackage[latin,english]{babel}
\usepackage{amsmath}
\usepackage{amssymb}
\usepackage{epstopdf}
\usepackage{graphics,psfrag,rotating}
\usepackage{mathabx}
\usepackage{graphicx}
\usepackage{dcolumn}
\usepackage{float}
\usepackage{pdflscape}
\usepackage{array}
\usepackage{booktabs}
\usepackage{amscd} 
\usepackage{mathtools}
\usepackage{fancybox}
\usepackage{fix-cm}
\usepackage[colorlinks=true,citecolor=blue,,linktocpage=true,linkcolor=blue,urlcolor=black]{hyperref}
\usepackage{tikz} 
\usetikzlibrary{matrix} 
\usetikzlibrary{positioning} 
\usetikzlibrary{arrows}
\usepackage{titlesec}
\usepackage{abstract}
\newcolumntype{M}[1]{>{\centering\arraybackslash}m{#1}}
\newcolumntype{N}{@{}m{0pt}@{}}

\def\be{\begin{equation}}
\def\ee{\end{equation}}
\def\bea{\begin{eqnarray}}
\def\eea{\end{eqnarray}}
\def\pd{\partial}
\def\a{\alpha}
\def\b{\beta}
\def\g{\gamma}
\def\d{\delta}
\def\m{\mu}
\def\n{\nu}
\def\t{\tau}

\def\l{\lambda}

\def\r{\rho}

\def\s{\sigma}

\def\bi{\begin{itemize}}
\def\ei{\end{itemize}}

\newcommand{\email}[1]{\href{mailto:#1}{\tt #1}}

\begin{document}

		\vspace*{-1cm}
		\phantom{hep-ph/***} 
		{\flushleft
			{{FTUAM-23-XXX}}
			\hfill{{ IFT-UAM/CSIC-23-15}}}
		\vskip 1.5cm
		\begin{center}
		{\LARGE\bfseries Physical charges versus conformal invariance
in unimodular gravity}\\[3mm]
			\vskip .3cm
		
		\end{center}
		\vskip 0.5  cm
		\begin{center}
			{\large Enrique Alvarez,
			Jesus Anero and Irene  Sanchez-Ruiz }
			\\
			\vskip .7cm
			{
				Departamento de F\'isica Te\'orica and Instituto de F\'{\i}sica Te\'orica, 
				IFT-UAM/CSIC,\\
				Universidad Aut\'onoma de Madrid, Cantoblanco, 28049, Madrid, Spain\\
				\vskip .1cm

				\vskip .5cm
				\begin{minipage}[l]{.9\textwidth}
					\begin{center} 
						\textit{E-mail:} 
						\email{enrique.alvarez@uam.es,jesusanero@gmail.com,irenesanchezl2@gmail.com}

					\end{center}
				\end{minipage}
			}
		\end{center}
	\thispagestyle{empty}
	
\begin{abstract}\vspace{-1em}
	\noindent
Unimodularity  can be implemented in different ways. In this paper we consider only the formulation of Unimodular Gravity in which the unimodular metric is obtained out of an unrestricted one as $\g_{\m\n}=|g|^{-{1\over n}} g_{\m\n}$. This procedure induces an extra Weyl symmetry. Some physical implications of this symmetry on the conserved currents are discussed.  Finally the results are illustrated  for the Painlev\'e-Gullstrand extension of Schwarzschild spacetime.
\end{abstract}

\newpage
\tableofcontents
	\thispagestyle{empty}
\flushbottom

\newpage
\setcounter{page}{1}
\section{Introduction}
Unimodular Gravity (UG) is a restriction of General Relativity to unimodular metrics
\be
g\equiv \left|\det\,g_{\m\n}\right|=1
\ee
A preliminary question is why Unimodular Gravity? The answer is that in this theory a constant vacuum density does not weigh, so that it does not induce a cosmological constant (cf. the first reference in \cite{AlvarezAnero}). Although that fact is not a full solution of the cosmological constant problem, it does solve half of it, namely why the vacuum energy usually believed to be generated in several phase transitions in the history of the universe do not induce a huge cosmological constant. It remains to explain the observed (small and positive) value of the cosmological constant. UG has nothing new to say about this.

In this paper, we want to study the conserved currents for the Unimodular Gravity  in two equivalent formulation of General Relativity (GR), the Hilbert Lagrangian
\be S_{\text{\tiny{H}}}[\g]\equiv -\frac{1}{2\kappa^2}\int
d^n x \hat{R}\ee
(our notation is that the metric $\g_{\m\n}$ is always unimodular and geometrical quantities constructed out of it are represented by hatted symbols) and the so-called Einstein Lagrangian\footnote{ We do not want here to take issues on the priority of Hilbert over Einstein in coming up with this Lagrangian. In fact this is usually (also by ourselves) called the Einstein-Hilbert Lagrangian; we have chosen the names in order to be able to tell apart two different candidates for the  GR Lagrangian.} where a surface term is added that eliminates the second derivatives of the metric. Staring from the identity\cite{Schrodinger}
\be \g^{\m\n}\hat{R}_{\m\n}=\partial_\r\left(\g^{\m\n}\hat{\Gamma}^\r_{\m\n}\right)-\partial_\n\left(\g^{\m\n}\hat{\Gamma}^\r_{\m\r}\right)+\g^{\m\n}\left(\hat{\Gamma}^\r_{\l\n}\hat{\Gamma}^\l_{\m\r}-\hat{\Gamma}^\r_{\l\r}\hat{\Gamma}^\l_{\m\n}\right)\ee
and defining Einstein's lagrangian from 
\be
 S_{\text{\tiny{E}}}\equiv \frac{1}{2\kappa^2}\int d^n x \g^{\m\n}\left(\hat{\Gamma}^\r_{\l\r}\hat{\Gamma}^\l_{\m\n}-\hat{\Gamma}^\r_{\l\n}\hat{\Gamma}^\l_{\m\r}\right)\ee

\be
 S_{\text{\tiny{H}}}=S_{\text{\tiny{E}}}+\int d^n x\,\pd_\m V^\m
 \ee
 with 
 \be
 V^\m={1\over 2 \kappa^2}\left(\hat{\Gamma}^\m_{\r\s}\g^{\r\s}-\g^{\m\n}\hat{\Gamma}_{\n\l}^\l\right)
 \ee
  that is,  Hilbert's  and Einstein's Lagrangians differ by a total derivative  $\pd_\m V^\m$ , so that both lagrangians  yield the same equations of motion.

The reason  for us to  consider the Einstein Lagrangian stems from the fact that it only involves first derivatives of the metric. In fact this is a well known way to  understand why Einstein equations are second order.
 A recent review about Unimodular Gravity can be found in \cite{AVA}.\\

There seems not to be any  free lunch in the universe, and the price to pay in this case is that the restriction to unimodular metrics implies that the theory is not invariant under the full diffeomorphism group (Diff), which is the invariance group of General Relativity, but only under volume preserving diffeomorphisms. This subgroup of the total diffeomorphism group, consists on those diffeomorphisms (connected with the identity)  that have unit Jacobian. 

\be
x\rightarrow x^\prime;  \quad \text{det}\,{\pd x^\prime\over \pd x}=1
\ee

The tangent space of the   {\em identity component} of volume preserving diffeomorphisms is generated by \footnote{
Diffeomorphism groups are rather tricky. For example \cite{Mann} there are diffeomorphisms arbitrarily close to the identity which can not  be obtained through the exponential map of the algebra.We shall in fact only consider in this work those diffeomorphisms that can be generated via the exponential map. We will always refer to this as the {\em algebra} of the group and do not worry about topological considerations.} {\em transverse vectors} $\xi_T^\m\in$ TDiff. 
\be\label{pd}
\hat{\nabla}_\m \xi_T^\m=\partial_\m\xi^\m_T=0
\ee
note, the covariant derivative is defined with the unimodular metric, $\g_{\m\n}$, therefore
\be \hat{\Gamma}^\l_{\l\m}=\frac{1}{\sqrt{\g}}\,\pd_\m\sqrt{\g}=0\ee

Once we have motivated our interest about Unimodular Gravity,  a new question appears, how we implement the unimodular constrait in the action. There are at least three ways to do it
\bi
\item The first one is through a Lagrange multiplier \cite{Kugo} i.e., add to the Lagrangian $\mathcal{L}$ the term
\be
S=\int d^nx\sqrt{|g|}\left[\mathcal{L}+\l(x)\left(g-1\right)\right]
\ee
This option works fine at the classical level, but does not work for quantum corrections \cite{Carmelo}.
\item Another option \cite{Eichhorn}, it using an exponential representation of a unimodular metric as the exponential of a traceless one
\be
g_{\m\n}(x)=e^{G_{\m\n}(x)}
\ee
where $\text{tr}\,G_{\m\n}(x)=0$.
\item The one we will use, which is to work with an unrestricted metric, $g_{\a\b}$ and construct the unimodular one through
\be
\g_{\a\b}=|g|^{-\frac{1}{n}}\,g_{\a\b}
\ee
This formalism has been used extensively in \cite{Alvarez:2015sba}.
\ei

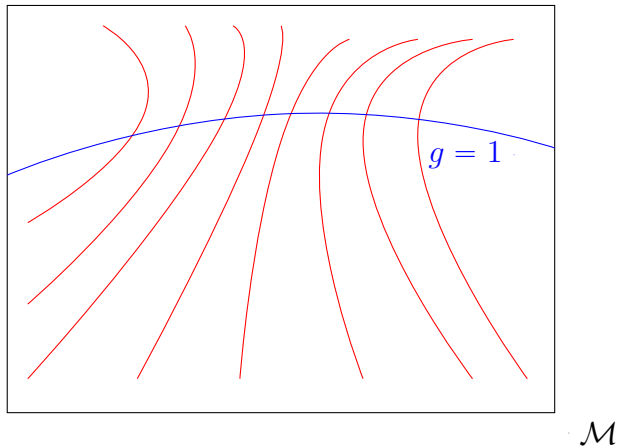
\begin{figure}[h!]\centering
	\begin{tikzpicture}[scale=1.8]
	\draw[black] (-1,-1) rectangle (3,2);
	\draw [red] plot [smooth, tension=1] coordinates {(-0.85,0.4) (0,1.2) (-0.3,1.85)};
	\draw [red] plot [smooth, tension=1] coordinates {(-0.85,-0.2) (0.2,1) (0.3,1.85)};
	\draw [red] plot [smooth, tension=1] coordinates {(-0.85,-0.75) (0.5,1) (0.65,1.85)};
	\draw [red] plot [smooth, tension=1] coordinates {(-0.05,-0.75) (0.8,1) (1,1.85)};
	\draw [red] plot [smooth, tension=1] coordinates {(0.7,-0.75) (1,1) (1.5,1.75)};
	\draw [red] plot [smooth, tension=1] coordinates {(1.6,-0.75) (1.3,1) (2,1.75)};
	\draw [red] plot [smooth, tension=1] coordinates {(2.4,-0.75) (1.6,1) (2.4,1.75)};
	\draw [red] plot [smooth, tension=1] coordinates {(2.8,-0.75) (2,1) (2.7,1.75)};
	\draw [blue] plot [smooth, tension=1] coordinates {(-1,0.75) (1,1.2) (3,0.95)};
	\filldraw [black] (3.1,-1.15) circle (0pt) node[anchor=west]{$\mathcal{M}$};
	\filldraw [blue] (2.7,0.9) circle (0pt) node[anchor=east]{$g=1$};
	\end{tikzpicture}   
	\caption{Weyl gauge orbits and gauge fixing for metrics $|\g|=1$. TDiff orbits are not parallel to Weyl orbits. } \label{figure}
\end{figure}

Recall that this third option  induces a (tautological) Weyl invariance
\be
g_{\a\b}\longrightarrow \Omega^2(x) g_{\a\b}
\ee
with $\Omega(x)=|g|^{-\frac{1}{2n}}$, therefore the total symmetry group of the theory is now
\begin{equation}
\text{WTDiff}= \text{Weyl}\rtimes \text{TDiff}
\end{equation}
Physically, what happens is that in the Weyl orbit of spacetimes that are gauge equivalent to the physical one, see the figure \ref{figure}, which by hypothesis  is the unimodular one, $\g_{\a\b}$ most naively defined physical characteristics are not constant (because they are not Weyl gauge invariant) (cf. figure). 

The plan of the paper is as follows. In the second section, we  briefly review about  Noether's current, and we make explicit our results in Hilbert's and Einstein's Lagrangian. We eventually find Weyl invariant currents, which reduce to General Relativity ones in the Weyl gauge fixed $g=1$. Related work on Noether currents in general Weyl invariant theories has recently appeared \cite{Alonso-Serrano} although the emphasis there is slightly different from ours. The section three, we shall apply the previous results for  the Painlev\'e-Gullstrand extension of Schwarzschild spacetime. Finally we present our conclusions.

The main goal of this paper is the conformally invariant expression of the physical currents in Unimodular Gravity.

\section{Conserved currents in Unimodular Gravity.}\label{N}
Let us  begin with a short review of the consequences of the  full Diff symmetry. After that, we study the special case of UG with only TDiff symmetry and  finally we derive our main results in the case of  Weyl$\rtimes$TDiff symmetry for Hilbert's and Einstein's Lagrangians.

Assume a Lagrangian, $\mathcal{L}$,    scalar under the full Diff group.  The invariance of the action under an arbitrary diffeomorphism generated by $\xi^\m$ implies that the variation of the action  reads
\be 
 \d S=\int d^nx\,\partial_\m (\mathcal{L}\xi^\m)
 \ee
On the other hand it is well-known that under a general variation of the metric (without changing the domain of integration) the variation of the action reads
\be
\d S=\int d^nx\sqrt{|g|} \bigg\{ \frac{\d S}{ \d g_{\a\b}} \d g_{\a\b} +\nabla_\m j^\m \bigg\}
\ee
on shell, that is, on the submanifold that obeys the equation of motion, this variation reduces to
\be
\d S=\int d^nx\sqrt{|g|} \nabla_\m j^\m=\int d^nx \partial_\m (\sqrt{|g|}j^\m)
\ee
The fact that this two variations have to be consistent independently of the integration domain means that the Noether current
\be \partial_\m j^\m_N=\partial_\m(\mathcal{L}\xi^\m-\sqrt{g}j^\m)=0\ee
has got to be conserved.

Let us derive  Noether's  current for a general Lagrangian $\mathcal{L}(g_{\a\b},\partial_\m g_{\a\b}, \partial_\m\partial_\n g_{\a\b})$. The variation of Lagrangian is
\be 
\d \mathcal{L}=\frac{\partial \mathcal{L}}{\partial g_{\a\b}}\d g_{\a\b}+\frac{\partial \mathcal{L}}{\partial \partial_\m g_{\a\b}}\d\partial_\m g_{\a\b}+\frac{\partial \mathcal{L}}{\partial \partial_\m\partial_\n g_{\a\b}}\d \partial_\m\partial_\n g_{\a\b}
\ee
Euler-Lagrange's equation reads
\be \frac{\partial \mathcal{L}}{\partial g_{\a\b}}-\partial_\m \frac{\partial \mathcal{L}}{\partial \partial_\m g_{\a\b}}+\partial_\m\partial_\n \frac{\partial \mathcal{L}}{\partial \partial_\m\partial_\n  g_{\a\b}}=0\ee
and Noether's current is
\be 
j_N^\m=\Big(\frac{\partial \mathcal{L}}{\partial \partial_\m g_{\a\b}}\d g_{\a\b}-\partial_\n\Big(\frac{\partial \mathcal{L}}{\partial \partial_\m\partial_\n g_{\a\b}}\Big)\d g_{\a\b}+\frac{\partial \mathcal{L}}{\partial \partial_\m\partial_\n g_{\a\b}}\partial_\n \d g_{\a\b}
\ee
which for a Diff generated by the vector field $\xi^\m$ can be written as

\be\label{jN} j_N^\m=\sqrt{|g|}\left( T_\r^{~\m}\xi^{\r}+U_\r^{~\m\n}\partial_\n\xi^{\r}+V_\r^{~\m(\l\t)}\partial_\l\partial_\t \xi^{\r}\right)\ee
with
\bea
T_\r^{~\m}&&=\d^\m_\r\mathcal{L}-\Big(\frac{\partial \mathcal{L}}{\partial \partial_\m g_{\a\b}}-\partial_\n\frac{\partial \mathcal{L}}{\partial \partial_\m\partial_\n g_{\a\b}}\Big)\partial_\r g_{\a\b}-\frac{\partial \mathcal{L}}{\partial \partial_\m\partial_\n g_{\a\b}}\partial_\n\partial_\r g_{\a\b}\nonumber\\
U_\r^{~\m\n}&&=2\Big(\partial_\t\frac{\partial \mathcal{L}}{\partial \partial_\m\partial_\t g_{\a\b}}-\frac{\partial \mathcal{L}}{\partial \partial_\m g_{\a\b}}\Big)g_{\r\s}I^{\n\s}_{\a\b}-\frac{\partial \mathcal{L}}{\partial \partial_\m\partial_\n g_{\a\b}}\partial_\r g_{\a\b}-2\frac{\partial \mathcal{L}}{\partial \partial_\m\partial_\t g_{\a\b}}I^{\n\s}_{\a\b}\partial_\t g_{\r\s}\nonumber\\
V_\r^{~\m(\l\t)}&&=-\frac{\partial \mathcal{L}}{\partial \partial_\m\partial_\t g_{\a\b}}g_{\r\s}I^{\l\s}_{\a\b}-\frac{\partial \mathcal{L}}{\partial \partial_\m\partial_\l g_{\a\b}}g_{\r\s}I^{\t\s}_{\a\b}
\eea
with
\be I^{\m\n}_{\a\b}=\frac{1}{2}\Big(\d^\m_\a\d^\n_\b+\d^\n_\a\d^\m_\b\Big)\ee

What changes in UG, where we only have TDiff invariance? 
The fact that the gauge parameters are not independent is the source of some small technical complications. One way out is to write
 \be
 \xi_T^\m=\nabla_\l\Omega^{\l\m},
 \ee
  where the tensor
\be \Omega_{\m\n}=-\Omega_{\n\m}\ee
is antisymmetric.
\par
Let us define in general the equations of motion as 
\be
H_{\m\n}\equiv {\d S\over \d g^{\m\n}}
\ee
TDiff  invariance \cite{AlvarezQG} implies that 
\be
\d S= \int d^nx\sqrt{|g|} H_{\m\n} \left(\nabla^\m \xi^\n+\nabla^\n \xi^\m\right)=\int d^nx\sqrt{|g|} H_{\m\n}\left(\nabla^\m  \nabla_\a \Omega^{\a\n}+\nabla^\n \nabla_\a \Omega^{\a\m}\right)
\ee
assuming that $\Omega_{\a\b}$ vanishes at the boundary,  $\left.\Omega_{\a\b}\right|_{\pd M}=0$, we can integrate by parts  \cite{Alvarez}
\be
\d S= \int d^nx\sqrt{|g|}\Big[\nabla_\a\nabla^\l H_{\l\b}-\nabla_\b\nabla^\l H_{\l\a}\Big]\Omega^{\a\b}
\ee
therefore\footnote{
	When there is ambiguity, we follow the rule of underlining the indices which undergo antisymmetrization.
}
\be
\nabla_{[\underline{\a}} \nabla^\l H_{\l\underline{\b}]}=0\ee

In conclusion, the Noether current \eqref{jN}, in function of the $\Omega_{\a\b}$ reads
\be j^\m_N=\sqrt{|g|}\left(T^\m_\r\nabla_\a\Omega^{\a\r}+U_{\r}^{\m\n}\nabla_\n\nabla_\a\Omega^{\a\r}+V_\r^{\m(\l\t)}\nabla_\l\nabla_\t \nabla_\a\Omega^{\a\r}\right)\ee
and it satisfies
\bea &&0=\int d^n x\partial_\m \Big[\sqrt{g}\left(T^\m_\r\nabla_\a\Omega^{\a\r}+U_{\r}^{\m\n}\nabla_\n\nabla_\a\Omega^{\a\r}+V_\r^{\m(\l\t)}\nabla_\l\nabla_\t \nabla_\a\Omega^{\a\r}\right)\Big]\eea
Given our assumption on the behavior of $\Omega^{\m\n}$, we can integrate by parts  
\bea
&&0=\int d^nx \sqrt{g}\Big[-(\nabla_\a \nabla_\m T^\m_\r)\Omega^{\a\r}+( T^\m_\r+\nabla_\l U_{\r}^{\l\m})\nabla_\m\nabla_\a\Omega^{\a\r}+\nonumber\\
&&+( U_{\r}^{\m\n}+\nabla_\l V_\r^{\l(\m\n)})\nabla_\m\nabla_\n\nabla_\a\Omega^{\a\r}+ V_\r^{\m(\l\t)}\nabla_\m\nabla_\l\nabla_\t \nabla_\a\Omega^{\a\r}\Big]\eea
and finally
\bea\label{cT1}
&\nabla_{[\underline{\a}} \nabla_\m T^\m_{\underline{\r}]}+\nabla_{[\underline{\a}} \nabla_\n \nabla_\m U_{\underline{\r}]}^{\m\n}+\nabla_{[\underline{\a}} \nabla_\m\nabla_\n \nabla_\l V_{\underline{\r}]}^{\l(\m\n)}=0
\eea
which is not a conservacion law as such, but rather a constraint on a covariant derivative of a multi-index current.
As has been already said,  Weyl's invariance of Unimodular Gravity is, in some sense, {\em tautological} (this is what Duff \cite{Duff} calls {\em pseudo-Weyl invariance}), in the sense that there is a field redefinition where the symmetry disappears.

We end this section with our main result, namely   Noether's current for the Hilbert and Einstein Lagrangian, 
(The same Lagrangians are studied in the paper \cite{Julia:1998ys} in the GR case).
\subsection{  Hilbert's UG   Lagrangian}
The Unimodular Hilbert action is 
\be S_{\text{\tiny{H}}}[\g]\equiv -\frac{1}{2\kappa^2}\int
d^n x \hat{R}\ee
where  the metric is unimodular $\g=|\text{det} \g_{\m\n}|=1$. As has been already remarked, it is possible to formulate the theory in such a way that it has an added Weyl invariance by writing
\be \g_{\m\n}=|g|^{-\frac{1}{n}}g_{\m\n}\ee
if we apply this transformation, the action for this new unrestricted metric reads
\be \label{UGH}S_{\text{\tiny{H}}}[g]=-\frac{1}{2\kappa^2}\int
d^n x g^{\frac{1}{n}}\Big[R-\frac{(n-1)(5n-2)}{4n^2}g^{\m\n}\frac{\nabla_\m g\nabla_\n g}{g^2}+\frac{(n-1)}{n}\frac{\Box g}{g}\Big]\ee

Under a general variation   $\d g_{\m\n}\equiv h_{\m\n}$ (forgetting the prefactor $-{1\over 2 \kappa^2}$ but keeping  total derivatives)
\be
\d S_{\text{\tiny{H}}}=\int d^nx\sqrt{|g|} \bigg\{ \frac{\d S}{ \d g_{\a\b}} \d g_{\a\b} +\nabla_\m j^\m \bigg\}
\ee
explicitly
\bea\label{EOMcSimply}
& \d S_{\text{\tiny{H}}}=g^{1/n}\Bigg\{\frac{1}{n}Rg_{\m\n}-R_{\m\n}+\frac{(n-2)(2n-1)}{4n^2}\left(\frac{\nabla_\m g\nabla_\n g}{g^2}-\frac{1}{n}\frac{\nabla_\a g\nabla^\a g}{g^2}g_{\m\n}\right)-\frac{n-2}{2n}\left(\frac{\nabla_\m\nabla_\n g}{g}-\frac{1}{n}\frac{\Box g}{g}g_{\m\n}\right)\Bigg\}\, h^{\m\n}+\nonumber\\
&+\sqrt{g}\Bigg\{\nabla_\m\Big[g^{\frac{2-n}{2n}}\nabla_\n h^{\m\n}\Big]-\frac{1}{2}\nabla_\n\Big[ h^{\m\n}g^{\frac{2-n}{2n}}\frac{\nabla_\m g}{g}\Big]-\frac{1}{n}\left(\nabla_\l\Big[g^{\frac{2-n}{2n}}\nabla^\l h\Big]-\frac{1}{2}\nabla_\l\Big[ h g^{\frac{2-n}{2n}}\frac{\nabla^\l g}{g} \Big]\right)\Bigg\}
\eea
The corresponding equation of motion is Weyl invariant and, up to total derivatives, traceless.
This means that
\bea\label{jH} j^\m&&=g^{\frac{2-n}{2n}}\Big[\nabla_\n h^{\m\n}-\frac{1}{2} h^{\m\n}\frac{\nabla_\n g}{g}-\frac{1}{n}\left(\nabla^\m h-\frac{1}{2} h \frac{\nabla^\m g}{g}\right)\Big]\eea
and using the previous results, the total TDiff  current  reads
\be j^\m_N=\sqrt{g}\left(T^\m_\r\nabla_\a\Omega^{\a\r}+U_{\r}^{\m\n}\nabla_\n\nabla_\a\Omega^{\a\r}+V_\r^{\m(\l\t)}\nabla_\l\nabla_\t \nabla_\a\Omega^{\a\r}\right)\ee
where
\bea\label{T1}
T^\m_\r&&=\d^\m_\r g^{\frac{2-n}{2n}}\Big[R-\frac{(n-1)(5n-2)}{4n^2}g^{\a\b}\frac{\nabla_\a g\nabla_\b g}{g^2}+\frac{(n-1)}{n}\frac{\Box g}{g}\Big]\nonumber\\
U_{\r}^{\m\n}&&=\frac{1}{2}g^{\frac{2-3n}{2n}}\Big[g^{\m\n}\nabla_\r g+\d^\m_\r\nabla^\n g-\frac{1}{n}\d^\n_\r\nabla^\m g\Big]\nonumber\\
V_\r^{\m(\l\t)}&&=g^{\frac{2-n}{2n}}\Big[-g^{\l\t}\d^\m_\r-\frac{n-2}{2n}\left(\d^\l_\r g^{\m\t}+\d^\t_\r g^{\m\l}\right)\Big]
\eea
In the unimodular Weyl gauge
\be g_{\m\n}\rightarrow\g_{\m\n} \hspace{0.5cm}\text{with}\hspace{0.5cm}\g=|\text{det} \g_{\m\n}|=1\ee
these expressions reduce to
\bea\label{T1}
\hat{T}^\m_\r&&=\hat{R} \d^\m_\r \nonumber\\
\hat{U}_{\r}^{\m\n}&&=0\nonumber\\
\hat{V}_\r^{\m(\l\t)}&&=-\g^{\l\t}\d^\m_\r-\frac{n-2}{2n}\left(\d^\l_\r \g^{\m\t}+\d^\t_\r \g^{\m\l}\right)
\eea
we continue to denote the unimodular objects with hatted symbols.

As a trivial check, under a Weyl transformation
\be \d g_{\a\b}=h_{\a\b}=2\omega g_{\a\b}\ee
the Weyl current  \eqref{jH} yields
\bea j_W^\m&&=g^{\frac{2-n}{2n}}\Big[2\nabla_\n(g^{\m\n}\omega) -\omega g^{\m\n}\frac{\nabla_\m g}{g}-2\nabla^\m \omega+\omega\frac{\nabla^\m g}{g}\Big]=0
\eea
which vanishes identically.  This is of course due to the tautological character of Weyl symmetry in our problem. 
\subsection{ Einstein's UG Lagrangian}
The Unimodular  Einstein  action \cite{Pauli} reads 
\be 
S_{\text{\tiny{E}}}\equiv \frac{1}{2\kappa^2}\int d^n x \g^{\m\n}\left(\hat{\Gamma}^\r_{\l\r}\hat{\Gamma}^\l_{\m\n}-\hat{\Gamma}^\r_{\l\n}\hat{\Gamma}^\l_{\m\r}\right)
\ee
In terms of the unrestricted metric 
\bea \label{UGE} S_{\text{\tiny{E}}}&&=\frac{1}{2\kappa^2}\int d^n x g^{1/n}\Big[ g^{\m\n}\left(\Gamma^\r_{\l\r}\Gamma^\l_{\m\n}-\Gamma^\r_{\l\n}\Gamma^\l_{\m\r}\right)-\frac{(n-2)}{2n}\left(g^{\m\n}\Gamma^\l_{\m\n}\frac{\nabla_\l g}{g}-\Gamma^\l_{\l\t}\frac{\nabla^\t g}{g}\right)-\nonumber\\
&&-\frac{(n-1)(n-2)}{4n^2}\frac{\nabla_\l g\nabla^\l g}{g^2}\Big]\eea
Under a general variation   $\d g_{\m\n}\equiv h_{\m\n}$ (keeping  total derivatives)
\be
\d S_{\text{\tiny{H}}}=\int d^nx\sqrt{|g|} \bigg\{ \frac{\d S}{ \d g_{\a\b}} \d g_{\a\b} +\nabla_\m j^\m \bigg\}
\ee
to be specific
\bea
&&\d S_{E}=g^{1/n}\Bigg\{\frac{1}{n}g_{\m\n}g^{\a\b}\left(\Gamma^\r_{\l\r}\Gamma^\l_{\a\b}-\Gamma^\r_{\l\a}\Gamma^\l_{\b\r}\right)-\left(\Gamma^\r_{\l\r}\Gamma^\l_{\m\n}-\Gamma^\r_{\l\n}\Gamma^\l_{\m\r}\right)+\nabla_\l\Gamma^\l_{\m\n}-\frac{1}{n}g_{\m\n}g^{\a\b}\nabla_\l\Gamma^\l_{\a\b}-\nonumber\\
&&-\nabla_\n \Gamma^\l_{\l\m}+\frac{1}{n}g_{\m\n}\nabla^\t \Gamma^\l_{\l\t}+\frac{(n-2)}{2n}\Big(\frac{\nabla_\m\nabla_\n g}{g}-\frac{1}{n}g_{\m\n}\frac{\Box g}{g}\Big)-\nonumber\\
&&-\frac{(n-2)(2n-1)}{4n^2}\Big( \frac{\nabla_\m g\nabla_\n g}{g^2}-\frac{1}{n}g_{\m\n}\frac{\nabla_\l g\nabla^\l g }{g^2}\Big)
\Bigg\}h^{\m\n}+\nonumber\\
&&+\sqrt{|g|}\Bigg\{-\nabla_\l\Big[g^{1/n-1/2}\Big(\Gamma^\l_{\m\n}h^{\m\n}-\frac{1}{n}g^{\a\b}\Gamma^\l_{\a\b}h\Big)\Big]+\nabla_\n\Big[g^{1/n-1/2}\Big(\Gamma^\l_{\l\m}h^{\m\n}-\frac{1}{n}g^{\m\n}\Gamma^\l_{\l\m}h\Big)\Big]-\nonumber\\
&&-\frac{(n-2)}{2n}\nabla_\n\Big[g^{1/n-3/2}\Big(h^{\m\n}\nabla_\m g-\frac{1}{n}hg^{\m\n}\nabla_\m g\Big)\Big]\Bigg\}
\eea
the corresponding equation  is Weyl invariant and, up to total derivatives, traceless. The total derivative term is generated by
\bea\label{jE} j^\m&&=-g^{1/n-1/2}\Big(\Gamma^\m_{\a\b}h^{\a\b}-\frac{1}{n}g^{\a\b}\Gamma^\m_{\a\b}h\Big)+g^{1/n-1/2}\Big(\Gamma^\l_{\l\n}h^{\m\n}-\frac{1}{n}g^{\m\n}\Gamma^\l_{\l\n}h\Big)-\nonumber\\
&&-\frac{n-2}{2n}g^{1/n-3/2}\Big(h^{\m\n}\nabla_\n g-\frac{1}{n}hg^{\m\n}\nabla_\n g\Big)\eea
and the Noether current is given by
\be j^\m_N=\sqrt{g}\left(T^\m_\r\nabla_\a\Omega^{\a\r}+U_{\r}^{\m\n}\nabla_\n\nabla_\a\Omega^{\a\r}+V_\r^{\m(\l\t)}\nabla_\l\nabla_\t \nabla_\a\Omega^{\a\r}\right)\ee
with
\bea\label{T2}
T^\m_\r&&=\d^\m_\r g^{\frac{2-n}{2n}}\Big[ g^{\a\b}\left(\Gamma^\t_{\l\t}\Gamma^\l_{\a\b}-\Gamma^\t_{\l\b}\Gamma^\l_{\a\t}\right)-\frac{(n-2)}{2n}\left(g^{\a\b}\Gamma^\l_{\a\b}\frac{\nabla_\l g}{g}-\Gamma^\l_{\l\t}\frac{\nabla^\t g}{g}\right)-\nonumber\\
&&-\frac{(n-1)(n-2)}{4n^2}\frac{\nabla_\l g\nabla^\l g}{g^2}\Big]\nonumber\\
U_{\r}^{\m\n}&&=g^{\frac{2-n}{2n}}\Big[2\,\Gamma^\m_{\a\r}g^{\a\n}-\frac{2}{n}g^{\a\b}\Gamma^\m_{\a\b}\d^\n_\r-\Gamma^\l_{\l\r}g^{\m\n}-\Gamma^\l_{\l\a}g^{\a\n}\d^\m_\r+\nonumber\\
&&+\frac{2}{n}g^{\m\t}\Gamma^\l_{\l\t}\d^\n_\r-\frac{n-2}{2n}\left(g^{\m\n}\frac{\nabla_\r g}{g}+\d^\m_\r\frac{\nabla^\n g}{g}-\frac{2}{n}\d^\n_\r \frac{\nabla^\m g}{g}\right)\Big]\nonumber\\
V_\r^{\m(\l\t)}&&=0
\eea
In the Weyl  unimodular gauge these expressions reduce to
\bea
\hat{T}^\m_\r&&=-\d^\m_\r  \g^{\a\b}\hat{\Gamma}^\t_{\l\b}\hat{\Gamma}^\l_{\a\t}\nonumber\\
\hat{U}_{\r}^{\m\n}&&=2\hat{\Gamma}^\m_{\a\r}\g^{\a\n}-\frac{2}{n}\g^{\a\b}\hat{\Gamma}^\m_{\a\b}\d^\n_\r\nonumber\\
\hat{V}_\r^{\m(\l\t)}&&=0
\eea

Under a Weyl rescaling also in this case Weyl's current 
\be
j_W^\m=0
\ee
vanishes identically. It is interesting that this result holds true in spite of the fact that Einstein's and Hilbert's lagrangians differ by a total derivative. Again, this is true due to the tautological character of Weyl's symmetry in this problem.
\par
Incidentally, the canonical energy-momentum tensor associated to $\mathcal{L}_E$ is precisely {\em Einstein's energy-momentum pseudotensor}. 
\be
t^{\text{\tiny{E}}}_{\m\n}\equiv \frac{\partial \mathcal{L}}{\partial(\partial_{\mu}g_{\alpha\b})}\partial^{\nu}g_{\alpha\b} - g^{\mu\nu}\mathcal{L}
\ee
and it is conserved in the sense that
\be
\pd_\m t^{\m\n}_{\text{\tiny{E}}}=0
\ee
in detail
\bea t_E^{\mu\nu} &&\equiv g^{1/n}\Big[\Gamma^\l_{\a\b}\Gamma^\t_{\t\l}g^{\a\m}g^{\b\n}-\Gamma^\l_{\l\a}\Gamma^\t_{\t\b}g^{\a\m}g^{\b\n}+\frac{1}{2}\Gamma^\l_{\l\b}\Gamma^\m_{\r\s}g^{\r\s}g^{\b\n}+\frac{1}{2}\Gamma^\l_{\l\b}\Gamma^\n_{\r\s}g^{\r\s}g^{\b\m}-\nonumber\\
&&-\Gamma^\r_{\b\l}\Gamma^\m_{\r\t}g^{\l\t}g^{\b\n}-\Gamma^\r_{\b\l}\Gamma^\n_{\r\t}g^{\l\t}g^{\b\m}+\frac{1}{2}\Gamma^\r_{\r\l}\Gamma^\m_{\b\t}g^{\l\t}g^{\b\n}+\frac{1}{2}\Gamma^\r_{\r\l}\Gamma^\n_{\b\t}g^{\l\t}g^{\b\m}-\nonumber\\
&&-\frac{(n-2)}{2n}\Big(\Gamma^\l_{\a\b}g^{\a\m}g^{\b\n}\frac{\nabla_\l g}{g}+\frac{1}{2}\Gamma^\m_{\b\l}g^{\b\n}\frac{\nabla^\l g}{g}+\frac{1}{2}\Gamma^\n_{\b\l}g^{\b\m}\frac{\nabla^\l g}{g}-\nonumber\\
&&-\Gamma^\l_{\b\l}g^{\b\n}\frac{\nabla^\m g}{g}-\Gamma^\l_{\b\l}g^{\b\m}\frac{\nabla^\n g}{g}\Big)\Big]- g^{\mu\nu}\mathcal{L}\eea
Finally, in the unimodular Weyl gauge, these expressions reduce to
\be
\hat{t}_E^{\mu\nu}=-\hat{\Gamma}^\r_{\b\l}\hat{\Gamma}^\m_{\r\t}\g^{\l\t}\g^{\b\n}-\hat{\Gamma}^\r_{\b\l}\hat{\Gamma}^\n_{\r\t}\g^{\l\t}\g^{\b\m}+\g^{\m\n}\g^{\a\b}\hat{\Gamma}^\r_{\l\a}\hat{\Gamma}^\l_{\b\r}
\ee
this result coincide  with the classical one in the Schr\"odinger book \cite{Schrodinger:2011gqa}.

We have studied the conserved current to Hilbert and Einstein Lagrangian, next, we want to apply our results in a particular case, the Schwarzschild metric.

\section{ Schwarzschild's UG  spacetime.}
As a simple application of the above, let us work out an explicit example, namely Schwarzschild spacetime from the Unimodular Gravity viewpoint. Schwarschild spacetime is still a solution of the equation of motion for Unimodular Gravity, although more general spherically symmetric vacuum solutions are possible. The reason is that for the usual Unimodular Hilbert Lagrangian
\be S_{\text{\tiny{H}}}[\g]\equiv -\frac{1}{2\kappa^2}\int
d^n x \hat{R}\ee
the equation of motion \eqref{EOMcSimply}, in the Weyl unitary gauge $\g=1$, reduces to the simple form of
\be
\hat{R}_{\m\n}-{1\over n} \hat{R} \g_{\m\n}=0
\ee
allowing Schwarzschild-(anti)-de Sitter spacetimes, in such a way that the naive version of Birkhoff's theorem is not valid.

We started with a particular set of coordinates for the Schwarzschild metric, the Painlev\'e-Gullstrand (PG) coordinates \cite{PG} 
\be ds^2=\left(1-\frac{r_s}{r} \right)\, dt^2\mp 2\sqrt{\frac{r_s}{r}} dt dr - dr^2-r^2\Big[\frac{d\psi^2}{1-\psi^2}+(1-\psi^2)d\phi^2\Big]\ee
where $r_s=2GM$. In fact  in \cite{Kraus} it was shown that with the upper sign this coordinate system covers regions I and II in Kruskal's diagram; whereas with the lower sign it covers regions $I^\prime$ and $II^\prime$. This is worked out in detail in \cite{Lemos}.

This metric  can be easily written in UG by means of a new coordinate $x$ such that the function $r(x)$, such that
\be r^4[x] (r'[x])^2=1\ee
which general solution is 
\be r=(3x+b)^{1/3}\ee
then the Painlev\'e-Gullstrand metric in unimodular gauge  reads
\be\label{metricU} ds^2=\left(1-\frac{r_s}{(3x+b)^{1/3}} \right)\, dt^2- \frac{2\sqrt{r_s}}{(3x+b)^{5/6}} dt dx - \frac{1}{(3x+b)^{4/3}}dx^2-(3x+b)^{2/3}\Big[\frac{d\psi^2}{1-\psi^2}+(1-\psi^2)d\phi^2\Big]\ee
The role of the constant $b$ is rather curious. As explained in \cite{Fromholz} Schwarzschild chose $b=r_s^3$, in order that the metric be regular everywhere except for a  singularity at $x=0$. This contributed to the initial confusion on the {\em Schwarzschild singularity.}

In spite of the fact that they are one of our main results we have relegated the explicit formulae to the appendix, owing to their unwieldly character.

\section{Conclusions}

In the Weyl invariant formulation of Unimodular Gravity the unimodular metric is represented as
\be
\g_{\a\b}\equiv g^{-\frac{1}{4}}\, g_{\a\b}
\ee
where the metric $g_{\a\b}$ is unrestricted. This introduces a Weyl abelian symmetry under
\be
g_{\a\b}\rightarrow \Omega^2(x)\,g_{\a\b}
\ee
The manifold of solutions is then a set of Weyl gauge orbits, with one representative in each orbit with a unimodular metric $\g_{\a\b}$ and the other metrics in the orbit related to that one by Weyl rescalings
\be
g_{\a\b}= g(x)^{1\over n} \g_{\a\b}
\ee
(with $g(x)$ an arbitrary function). 
\par
This Weyl invariance is a bit peculiar; we call it {\em tautological}, in the sense that it disappears under a field redefinition. This translates into the fact that the associated Weyl current vanishes identically (in Hilbert's version)
\be
j_W^\m=0
\ee
Let us make a comment on Weyl invariance. The horizon $r=r_S$ in Schwarzschild's spacetime has an associated Bekenstein-Hawking entropy \cite{Bekenstein} given by
\be
S={A\over 4 G}=\pi{r_s^2\over G}
\ee
which according to Wald \cite{Wald} can be interpreted as the value of the Noether charge associated to diffeomorphism invariance. In the Weyl orbit there are static spacetimes without a Killing horizon (of course there are also non-static spacetimes without any Killing).
Given the fact that in the vicinity of the horizon
\be
g_{00}\sim\frac{x-x_s}{r_s^3}+\ldots
\ee
(with $x_s\equiv \frac{r_s^3-b}{3}$)  the rescaling  
\be
g(x)\sim\frac{1}{(x-x_s)^4}
\ee
is singular at $x=x_S$, and there are corresponding singularities in the derivatives of $g(x)$. This is the the price to pay to get rid of the horizon. If only nonsingular rescalings are accepted, then  $g_{00}=0$ all along Weyl's orbit.
\par

\par
The issue treated in this paper is that physical characteristics of different spacetimes (like total mass in the asymptotically flat case) are not Weyl invariant. This means that in some sense naive TDiff conserved currents are not constant on Weyl orbits. It seems desirable to have a definition of TDiff conserved currents that is Weyl invariant. 
\par
What we provide in this paper is precisely such a Weyl gauge invariant formulation of the TDiff currents, although we claim that the physical value is the one corresponding to the unimodular gauge fixing. This is more or less obvious owing to the way that Weyl invariance has been introduced in the theory.


\section*{Acknowledgments}
 This work has received funding from the Spanish Research Agency (Agencia Estatal de Investigacion) through the grant IFT Centro de Excelencia Severo Ochoa MCIN CEX2020-001007-S.  This project has also received funding from the European UnionÕs Horizon 2020 research and innovation programme under the Marie Sklodowska -Curie grant agreement No 860881-HIDDeN.

\newpage
\appendix
\section{ PG Hilbert's  current.}
In this case the total current under TDiff reads
\be j^\m_N=\sqrt{g}\left(T^\m_\r\nabla_\a\Omega^{\a\r}+U_{\r}^{\m\n}\nabla_\n\nabla_\a\Omega^{\a\r}+V_\r^{\m(\l\t)}\nabla_\l\nabla_\t \nabla_\a\Omega^{\a\r}\right)\ee
in the unimodular Weyl gauge reduces to
\bea\label{T1}
\hat{T}^\m_\r&&=0 \nonumber\\
\hat{U}_{\r}^{\m\n}&&=0\nonumber\\
\hat{V}_\r^{\m(\l\t)}&&=-\g^{\l\t}\d^\m_\r-\frac{n-2}{2n}\left(\d^\l_\r \g^{\m\t}+\d^\t_\r \g^{\m\l}\right)
\eea
owing to the fact that the scalar of curvature vanishes. In full detail, by components
\bea
V^{t(tt)}_{t}&&= -2 + \frac{2}{n} \nonumber\\
V^{t(tx)}_{t}&&= V^{t(xt)}_{t}= \frac{(n-2) \sqrt{r (b+3x)}}{n}\nonumber\\
V^{t(xx)}_{t}&&= \frac{-2(b+3x)\left[(b+3x)^{1/3}-r\right]+  n \Big[-1 - 3 r x  +3x(b+3x)^{1/3}+ b \left[(b+3x)^{1/3}-r\right]\Big]}{n}\nonumber\\
V^{t(\psi\psi)}_{t}&&=  \frac{-b n +2 (b+3x)^{1/3}(-1 + \psi^2) +  n \Big[-3 x -  (b+3x)^{1/3}(-1 + \psi^2)\Big]}{n(b+3x)}\nonumber\\
V^{t(\phi\phi)}_{t}&&= \frac{2 (b+3x)^{1/3}+ b n(1 -   \psi^2) -  n \Big[(b+3x)^{1/3}+ 3 x (-1 + \psi^2)\Big]}{n (b+3x)(-1 + \psi^2)}
\eea
\bea
V^{x(tt)}_{x}&&= -1 + \frac{2}{n} - 3 xr + 3 x (b+3x)^{1/3}+ b \Big[(b+3x)^{1/3}-r\Big]\nonumber\\
V^{x(tx)}_{x}&&= V^{x(xt)}_{x}= \frac{(n-2) \sqrt{r_{s} (b+3x)}}{n}\nonumber\\
V^{x(xx)}_{x}&&=\frac{2 (n-1) (b+3x) \Big[(b+3x)^{1/3}-r\Big]}{n}\nonumber\\
V^{x(\psi\psi)}_{x}&&=(b+3x) \Big[(b+3x)^{1/3}-r\Big] -  \frac{(n-2) (-1 + \psi^2)}{n (b+3x)^{2/3}}\nonumber\\
V^{x(\phi\phi)}_{x}&&= -3 rx + b \Big[(b+3x)^{1/3}-r\Big]  + \frac{3 x (b+3x) + \frac{n-2}{n (-1 + \psi^2)}}{(b+3x)^{2/3}}
\eea
\bea
V^{\psi(tt)}_{\psi}&&= - \frac{b (n-2) + 3 (n-2) x + n (b+3x)^{1/3} (-1 + \psi^2)}{n(b+3x)}\nonumber\\
V^{\psi(tx)}_{\psi}&&= V^{\psi(xt)}_{\psi}= \frac{(n-2) \sqrt{r_{s} (b+3x)}}{n}\nonumber\\
V^{\psi(xx)}_{\psi}&&= \frac{r + \frac{1}{n}(n-2)(b+3x)^{2} \Big[(b+3x)^{1/3}-r\Big] -  r\psi^2 -  \Big[(b+3x)^{1/3}-r\Big] (-1 + \psi^2)}{b+3x}\nonumber\\
V^{\psi(\psi\psi)}_{\psi}&&= - \frac{2 (n-1) (-1 + \psi^2)}{n (b+3x)^{2/3}}\nonumber\\
V^{\psi(\phi\phi)}_{\psi}&&= \frac{2 -  n (2 - 2 \psi^2 + \psi^4)}{n (b+3x)^{2/3}(-1 + \psi^2)}\eea
\bea
V^{\phi(tt)}_{\phi}&&=  \frac{-n (b+3x)^{1/3} + b (n-2) (-1 + \psi^2) + 3 (n-2) x (-1 + \psi^2)}{n (b+3x) (-1 + \psi^2)}\nonumber\\
V^{\phi(tx)}_{\phi}&&=V^{\phi(xt)}_{\phi}= \frac{(n-2) \sqrt{r_{s} (b+3x)}}{n}\nonumber\\
V^{\phi(xx)}_{\phi}&&=\frac{\frac{1}{n}(n-2) (b+3x)^{2}\Big[(b+3x)^{1/3}-r\Big] -  \frac{r}{-1 + \psi^2} + \frac{1}{-1 + \psi^2}\Big[(b+3x)^{1/3}-r\Big]}{b+3x}\nonumber\\
V^{\phi(\psi\psi)}_{\phi}&&= \frac{-1 - \frac{1}{n}(n-2) (-1 + \psi^2)^2}{(b+3x)^{2/3}(-1 + \psi^2)}\nonumber\\
V^{\phi(\phi\phi)}_{\phi}&&= \frac{-2 (n-1)}{n(b+3x)^{2/3} (-1 + \psi^2)}
\eea

\be
V^{t(\l\t)}_{x}=V^{x(\l\t)}_{t}= \sqrt{r(b+3x)}
\ee
\section{  PG Einstein's current.}
We are sorry for the  formulas of this paragraph, but we did not find a nice geometrical way to write them.
\bea
\hat{T}^\m_\r&&=-\d^\m_\r  \g^{\a\b}\hat{\Gamma}^\t_{\l\b}\hat{\Gamma}^\l_{\a\t}\nonumber\\
\hat{U}_{\r}^{\m\n}&&=2\hat{\Gamma}^\m_{\a\r}\g^{\a\n}-\frac{2}{n}\g^{\a\b}\hat{\Gamma}^\m_{\a\b}\d^\n_\r\nonumber\\
\hat{V}_\r^{\m(\l\t)}&&=0
\eea
in the Weyl unimodular gauge
\be \hat{T}^\m_\r=\begin{pmatrix}
	\frac{2r}{b+3x}&\frac{2\sqrt{r}\left[(b+3x)^{1/3}-r\right]}{\sqrt{b+3x}}&0&0\\
	\frac{2\sqrt{r}}{(b+3x)^{3/2}}&\frac{2\left[(b+3x)^{1/3}-r\right]}{(b+3x)}&0&0\\
	0&0&0&0\\
	0&0&0&0\end{pmatrix}
\ee
\bea && U_{\r}^{\m\n} = \text{by components}\nonumber\eea
\begin{align*}
	U^{tt}_{t} & = \frac{r^{3/2}}{n(b+3x)^{5/6}} & U^{xx}_{t} & = \frac{-\sqrt{r}\left[r+(n-1)(b+3x)^{1/3}\right]}{n (b+3x)^{5/6}}\nonumber\\
	U^{\psi\psi}_{t} & =  U^{\phi\phi}_{t} = \frac{2(n-1)\sqrt{r}}{n \sqrt{b+3x}}\nonumber\\
	U^{tx}_{t} & = -\frac{r}{n } \Big[-1+n+\frac{r}{(b+3x)^{1/3}}\Big]& U^{xt}_{t} & = \frac{r}{n(b+3x)^{4/3}}\nonumber\\
\end{align*}

\begin{align*}
	U^{tt}_{x} & = \frac{r\left[(n-5)(b+3x)+4r(b+3x)^{2/3}\right]}{n(b+3x)} \nonumber\\
	U^{xx}_{x} & = \frac{1}{n}\Big[(7-3n)r-\frac{4r^2}{(b+3x)^{1/3}}+4(n-1)(b+3x)^{1/3}\Big]\nonumber\\
	U^{\psi\psi}_{x} &= U^{\phi\phi}_{x} =\frac{2(n-1)\left[(b+3x)^{1/3}-r\right]}{n} \nonumber\\
	U^{tx}_{x} & = \frac{-4\sqrt{r}(b+3x)^{1/6}\left[(b+3x)^{1/3}-r\right]^2}{n} &U^{xt}_{x} & = \frac{4\sqrt{r}\left[r+(n-1)(b+3x)^{1/3}\right]}{n (b+3x)^{5/6}} \nonumber\\
\end{align*}

\begin{align*}
	U^{x\psi}_{\psi} & = \frac{2 (-1 + \psi^2)}{(b+3x)^{5/3}} &U^{\psi t}_{\psi}& = - \frac{2 \sqrt{r}}{\sqrt{b+3x}}&U^{\psi x}_{\psi}& = 2 \left[(b+3x)^{1/3}-r\right]  \nonumber\\
	U^{\psi\psi}_{\psi} & = U^{\phi\phi}_{\psi}= \frac{-2(n-1)\psi}{n (b+3x)^{2/3}} \nonumber\\
\end{align*}

\begin{align*}
	U^{x\phi}_{\phi} & = \frac{2}{(b+3x)^{5/3}(-1 + \psi^2)} &U^{\psi \phi}_{\psi}& = \frac{2 \psi}{(b+3x)^{2/3} (-1 + \psi^2)^2}&U^{\phi t}_{\phi}& = - \frac{2 \sqrt{r}}{\sqrt{b+3x}}\nonumber\\
	U^{\phi x}_{\phi} & = 2 \left[(b+3x)^{1/3}-r\right]  &U^{\phi\psi}_{\phi}& = \frac{2 \psi}{(b+3x)^{2/3} }\nonumber\\
\end{align*}

Finally, in the unimodular Weyl gauge, the energy-momentum pseudotensor 
\be
\hat{t}_{\text{\tiny{E}}}^{\mu\nu}=-\hat{\Gamma}^\r_{\b\l}\hat{\Gamma}^\m_{\r\t}\g^{\l\t}\g^{\b\n}-\hat{\Gamma}^\r_{\b\l}\hat{\Gamma}^\n_{\r\t}\g^{\l\t}\g^{\b\m}+\g^{\m\n}\g^{\a\b}\hat{\Gamma}^\r_{\l\a}\hat{\Gamma}^\l_{\b\r}
\ee
for the the Painlev\'e-Gullstrand metric conducts to
\be \hat{t}_{\text{\tiny{E}}}^{\mu\nu}=\begin{pmatrix}
	\hat{t}^{tt}_{\text{\tiny{E}}}&0&0&0\\
	0&\hat{t}^{rr}_{\text{\tiny{E}}}&\hat{t}^{r\psi}_{\text{\tiny{E}}}&0\\
	0&\hat{t}^{\psi r}_{\text{\tiny{E}}}&\hat{t}^{\psi\psi}_{\text{\tiny{E}}}&0\\
	0&0&0&\hat{t}^{\phi\phi}_{\text{\tiny{E}}}\end{pmatrix}
\ee
in full detail

\bea \hat{t}^{tt}_{\text{\tiny{E}}}&&=\frac{1}{2} \Bigg\{- \frac{r_s^2}{(1 + 3 x)^{ 4/3} }-  \frac{r_s^2}{(2 + 3 x)^2 \bigl((2 + 3 x)^{1/3} -  r_s\bigr)^2} +\frac{(3x+b)^{1/3}}{(3x+b)^{1/3}-r_s}f[x,\psi,r_s]\Bigg\}\nonumber\\
\hat{t}^{rr}_{\text{\tiny{E}}}&&=- \frac{r_s^2}{2 (1 + 3 x)^{4/3}} -  \frac{\bigl(8 + 12 x - 3 (2 + 3 x)^{2/3} r_s\bigr)^2}{2 (2 + 3 x)^{10/3} \bigl((2 + 3 x)^{1/3} -  r_s\bigr)^2}- \frac{2 \bigl(3 + 3 x -  (3 + 3 x)^{2/3} r_s\bigr)}{3 (1 + x) (-1 + \psi^2)} - \nonumber\\
&& -  \frac{2 \bigl((4 + 3 x)^{1/3} -  r_s\bigr) (-1 + \psi^2) }{(4 + 3 x)^{1/3}}-\frac{1}{2}(3x+b)\left[(3x+b)^{1/3}-r_s\right]f[x,\psi,r_s]\nonumber\\
\hat{t}^{\psi\psi}_{\text{\tiny{E}}}&&=- \frac{2}{(2 + 3 x)^2} + 2 \psi^2 -  \frac{2 \psi^2}{(-1 + \psi^2)^2} -  \frac{2 \bigl(3 + 3 x -  (3 + 3 x)^{2/3} r_s\bigr)}{3 (1 + x) (-1 + \psi^2)}+\frac{(-1+\psi^2)}{2(3x+b)^{2/3}}f[x,\psi,r_s]\nonumber\\
\hat{t}^{\phi\phi}_{\text{\tiny{E}}}&&=2 -  \frac{2}{(2 + 3 x)^2} -  \frac{2 \psi^2}{(-1 + \psi^2)^2} + \frac{2 r_s (-1 + \psi^2)}{(4 + 3 x)^{1/3}}+\frac{1}{2(3x+b)^{2/3}(-1+\psi^2)}f[x,\psi,r_s]\nonumber\\
\hat{t}^{r\psi}_{\text{\tiny{E}}}&&=\hat{t}^{\psi r}_{\text{\tiny{E}}}=\psi \Bigg\{-4 - 3 x + (4 + 3 x)^{2/3} r_s + \frac{3 + 3 x -  3^{2/3} (1 + x)^{2/3} r_s}{(-1 + \psi^2)^2} +\nonumber\\
&&+ \frac{1}{(3 + 3 x) (-1 + \psi^2)} -  \frac{1 -  \psi^2}{4 + 3 x}\Bigg\}\nonumber\\
\eea
where
\bea f[x,\psi,r_s]=
&&\frac{2}{(2 + 3 x)^2} +\frac{1}{ (3 + 3 x)^{2}} + \frac{1}{(4 + 3 x)^{2}} + \frac{3 r_s^2}{4 (1 + 3 x)^{4/3}} + \frac{r_s^2}{4 (2 + 3 x)^{4/3}} +\nonumber\\
&&+ \frac{r_s^2}{4 (1 + 3 x)^2 \bigl((1 + 3 x)^{1/3} -  r_s\bigr)^2}+ \frac{r_s^2}{4 (2 + 3 x)^2 \bigl((2 + 3 x)^{1/3} -  r_s\bigr)^2} + \nonumber \\ 
&& + \frac{\bigl(8 + 12 x - 3 (2 + 3 x)^{2/3} r_s\bigr)^2}{2 (2 + 3 x)^{10/3} \bigl((2 + 3 x)^{1/3} -  r_s\bigr)^2} - 4 \psi^2 + \frac{4 \psi^2}{(-1 + \psi^2)^2}+ \nonumber \\ 
&& + \frac{3 + 3 x -  3^{2/3} (1 + x)^{2/3} r_s}{(1 + x) (-1 + \psi^2)} + \frac{2 + 3 x -  (2 + 3 x)^{2/3} r_s}{(2 + 3 x) (-1 + \psi^2)} + \nonumber \\ 
&& + \frac{\bigl((2 + 3 x)^{1/3} -  r_s\bigr) (-1 + \psi^2)}{(2 + 3 x)^{1/3}} +\frac{3 \bigl((4 + 3 x)^{1/3} - r_s\bigr) (-1 + \psi^2)}{(4 + 3 x)^{1/3}}
\eea
It this formalism, just by construction, the values of the physical energy /mass 
\be
m=\int t^0_0 d x\wedge d\psi\wedge d\phi
\ee
are the same in the whole Weyl orbit generated by $\Omega(x)$ (that is, they are independent of the Weyl rescaling $g$).
\newpage


\end{document}